\documentclass[prd,nofootinbib,preprint,superscriptaddress]{revtex4}
\pdfoutput=1
\usepackage[T1]{fontenc}
\usepackage{amsmath,amssymb}
\usepackage{braket}
\usepackage{epsfig}
\usepackage{graphicx}
\usepackage[usenames,dvipsnames]{color}
\usepackage{caption}
\usepackage{subcaption}
\usepackage{slashed}
\usepackage{mathtools}
\usepackage[colorlinks,citecolor=blue]{hyperref}
\usepackage{pdfpages}
\usepackage{color}
\usepackage{xcolor}
\usepackage{comment}
\usepackage{tikz}
\usepackage{tikz-feynman}
\usetikzlibrary{positioning}
\usetikzlibrary{decorations.pathmorphing}

\newcounter{figfn}

\begin{document}



\title{Multi-peaked high-frequency gravitational waves from PBH-assisted leptogenesis}

\author{Debasish Borah}
\email{dborah@iitg.ac.in}
\affiliation{Department of Physics, Indian Institute of Technology Guwahati, Assam 781039, India}

\author{Nayan Das}
\email{nayan.das@iitg.ac.in}
\affiliation{Department of Physics, Indian Institute of Technology
Guwahati, Assam 781039, India}

\begin{abstract}
We study the possibility of probing non-thermal leptogenesis with multi-peaked high-frequency gravitational waves (GW) by considering heavy right-handed neutrino (RHN) produced from primordial black hole (PBH) evaporation to be responsible for generating the required lepton asymmetry. The decay of RHN also produces a GW spectrum due to graviton bremsstrahlung with the corresponding amplitude being enhanced for heavier RHN. The presence of an ultra-light PBH dominated epoch not only ensures sufficient production of RHNs, but also keeps the leptogenesis scenario free from strong washout problem of thermal leptogenesis at very high scale. In addition, the PBH dominated epoch also keeps the RHN-generated gravitational bremsstrahlung spectrum distinct from the stochastic GW background from the thermal bath. Finally, PBH evaporation also brings two separate sources of GW via density perturbation and graviton emission via Hawking evaporation. For the most optimistic scenario with very high scale seesaw consistent with neutrino mass and leptogenesis, this leads to a multi-peaked GW spectrum with peak frequencies lying in the MHz-EHz range.
\end{abstract}
\maketitle
\section{Introduction}
The visible matter content in the present Universe is highly asymmetric, referred to as the baryon asymmetry of the Universe (BAU) \cite{Planck:2018vyg, ParticleDataGroup:2024cfk}. Starting with a baryon symmetric Universe, such an asymmetry can be generated dynamically if Sakharov's conditions are satisfied \cite{Sakharov:1967dj}. However, the standard model (SM) of particle physics fails to satisfy them in required amount, making the origin of BAU a longstanding puzzle in particle physics and cosmology. This has led to several beyond standard model (BSM) frameworks like baryogenesis \cite{Weinberg:1979bt, Kolb:1979qa} as well as leptogenesis \cite{Fukugita:1986hr}. In leptogenesis, a non-zero lepton asymmetry is generated first which later gets converted into baryon asymmetry by electroweak sphalerons \cite{Kuzmin:1985mm}. One interesting aspect of leptogenesis is its connection to the origin of light neutrino masses within canonical seesaw mechanisms like type-I seesaw \cite{Minkowski:1977sc, Yanagida:1979as, Gell-Mann:1979ijt, GellMann:1980vs, Mohapatra:1979ia, Schechter:1980gr, Schechter:1981cv}. In such a scenario, the out-of-equilibrium CP violating decay of a heavy right handed neutrino (RHN) is responsible for generating the required lepton asymmetry. In order to generate non-zero CP asymmetry while being consistent with neutrino oscillation data, at least two copies of such RHNs are required. Additionally, with hierarchical RHN spectrum, there exists a lower bound on the scale of thermal leptogenesis, known as the Davidson-Ibarra (DI) bound $m_{N_{1}} \gtrsim 10^9$ GeV \cite{Davidson:2002qv}. This keeps canonical leptogenesis models out of reach from terrestrial experiments.

Due to the absence of direct signatures, there have been attempts to probe high scale seesaw or leptogenesis scenarios via stochastic gravitational wave (GW) observations \cite{Dror:2019syi, Blasi:2020wpy, Fornal:2020esl, Samanta:2020cdk, Barman:2022yos, Huang:2022vkf, Dasgupta:2022isg, Okada:2018xdh, Hasegawa:2019amx, Chianese:2025mll, Borah:2022cdx, Borah:2022vsu, Barman:2023fad, Borah:2023saq, Borah:2025bfa, Borah:2026kfo} with the GW source being cosmic strings \cite{Dror:2019syi, Blasi:2020wpy, Fornal:2020esl, Samanta:2020cdk, Borah:2022vsu}, domain walls \cite{Barman:2022yos, Barman:2023fad, Saikawa:2017hiv, Roshan:2024qnv, Bhattacharya:2023kws, Blasi:2022ayo, Blasi:2023sej, Borah:2024kfn, Paul:2024iie, Borboruah:2024lli, Borah:2025bfa, Borah:2026kfo} or bubbles generated at first order phase transition (FOPT) \cite{Huang:2022vkf, Dasgupta:2022isg, Okada:2018xdh, Hasegawa:2019amx, Borah:2022cdx, Borah:2023saq}. However, all these probes require extensions of the minimal seesaw model and the origin of GW is not directly connected to the origin of lepton asymmetry. Recently, there have been attempts to find GW signature of such leptogenesis scenarios directly from the decay of heavy RHN or similar fields involved in other seesaw-leptogenesis models \cite{Datta:2024tne, Choi:2025hqt, Murayama:2025thw, Kanemura:2025rct, Wang:2025mtq}. This involves gravitons emitted via bremsstrahlung-like process during RHN decay into lepton and Higgs. As pointed out in some of these works, the GW amplitude is enhanced with some hope for future detection only for very heavy RHN having mass close to the grand unified theory (GUT) scale $\sim 10^{16}$ GeV. For thermal leptogenesis, this requires a very high reheat temperature, saturating the upper bound from cosmic microwave background (CMB) observations \cite{Planck:2018jri}. On the other hand, thermal leptogenesis with very heavy RHN mass suffers from the strong washout problem due to fast $\Delta L=2$ processes \cite{Buchmuller:2004nz, Giudice:2003jh}. Finally, the peak of the GW spectrum from RHN decay remains buried under the GW background spectrum from thermal bath \cite{Ghiglieri:2015nfa, Ghiglieri:2020mhm, Ringwald:2020ist}, making it difficult for experimental verification of the contribution from RHN decay alone.

Motivated by these, we consider a non-thermal leptogenesis scenario to avoid the above-mentioned issues related to thermal leptogenesis. While this can occur due to non-thermal production of RHNs during reheating after inflation~\cite{Lazarides:1991wu, Murayama:1992ua, Kolb:1996jt, Giudice:1999fb, Asaka:1999yd, Asaka:1999jb, Hamaguchi:2001gw, Hahn-Woernle:2008tsk, Hamada:2018epb, Eijima:2019hey, Maleknejad:2020pec, Barman:2021tgt, Barman:2021ost, Barman:2022gjo, Lazarides:2022ezc, Datta:2022jic, Datta:2023pav, Barman:2024jqh, Barman:2024ujh, Chowdhury:2026zox}, we consider the production of RHN from the evaporation of ultra-light primordial black holes (PBH) \cite{Hawking:1974rv, Hawking:1975vcx,Carr:2020gox} with initial mass $M_\text{in}\lesssim 3.4\times 10^8\,\text{g}$ which evaporate before the epoch of the big bang nucleosynthesis (BBN). Graviton bremsstrahlung from RHNs produced from reheating has already been studied in \cite{Kanemura:2025rct}. While the authors of \cite{Choi:2025hqt} considered a setup with ultra-light PBH and type-I seesaw, they considered leptogenesis from thermal RHNs while graviton bremsstrahlung occurs dominantly from an ultra-long-lived RHN produced from PBH evaporation. Graviton bremsstrahlung from heavy scalar particle produced from PBH evaporation was studied by \cite{Choi:2024acs} but without any connection to the origin of neutrino mass or leptogenesis. In this work, we consider a type-I seesaw leptogenesis scenario with a PBH dominating epoch such that heavy RHNs are produced non-thermally and hence does not suffer from the problem of strong $\Delta L=2$ washout \cite{Bernal:2022pue}. Due to the early matter dominated phase of ultra-light PBH, it also separates the RHN decay originated GW spectra from the thermal background resulting in two distinct peaks. The PBH dominated epoch also sources other GW background PBH density fluctuations \cite{Papanikolaou:2020qtd, Domenech:2020ssp, Domenech:2021wkk, Domenech:2021ztg, Papanikolaou:2022chm} as well as Hawking evaporation into gravitons \cite{Anantua:2008am}. For the most optimistic scenario with the largest possible GW amplitude from RHN decay, we find a high-frequency GW spectrum with three distinct peaks with the chosen parameter space being consistent with the observed neutrino oscillation data and baryon asymmetry via leptogenesis.

This paper is organized as follows. In section \ref{sec1}, we summarize the basics of PBH-assisted leptogenesis. In section \ref{sec2}, we discuss different GW sources in this scenario and finally conclude in section \ref{sec3}


\section{PBH-assisted leptogenesis}
\label{sec1}

\subsection{Primordial Black Holes}
Primordial black holes, originally proposed in \cite{Zeldovich:1967lct} and later by Hawking \cite{Hawking:1974rv, Hawking:1975vcx}, have gained significant attention due to their cosmological consequences \cite{Chapline:1975ojl, Carr:1976zz}. A recent review of PBH can be found in \cite{Carr:2020gox}. In the early universe, PBH can be formed in a variety of ways like, from inflationary perturbations \cite{Hawking:1971ei, Carr:1974nx, Wang:2019kaf, Byrnes:2021jka, Braglia:2022phb}, first-order phase transition (FOPT) \cite{Crawford:1982yz, Hawking:1982ga, Moss:1994iq, Kodama:1982sf}, the collapse of topological defects \cite{Hawking:1987bn, Deng:2016vzb} etc. We remain agnostic about their specific production mechanism but assume their formation in a radiation-dominated epoch after the end of inflation. We also assume a monochromatic mass spectrum for PBH with initial mass $M_{\rm in}$, zero spin, and vanishing electric charge. Apart from initial mass, another key parameter related to PBH is the initial energy fraction 
\begin{eqnarray}
\beta \equiv \frac{\rho_{\rm BH}(T_{\rm in})}{\rho_{\rm R}(T_{\rm in})},
\end{eqnarray}
where $T_{\rm in}$ denotes the plasma temperature at the time of PBH formation, while $\rho_{\rm BH}$ and $\rho_{\rm R}$ represent the PBH and radiation energy densities, respectively.  {After their formation, the energy density of PBH scales as $a^{-3}$, whereas the radiation energy density redshifts as $a^{-4}$. As a result, PBH may dominate the energy density of the Universe for a brief period before completely evaporating. The condition for PBH domination is $\beta > \beta_{c}$, where $\beta_{c} \simeq 5.6 \times 10^{-6} \left(\frac{1 \text{g}}{M_{\rm in}}\right)$. }

For PBH formed during radiation domination, the initial PBH mass is typically a fraction of the horizon mass at formation and can be expressed as
\begin{equation}
M_{\rm in}
=
\gamma
\frac{4\pi}{3\,\mathcal{H}^3(T_{\rm in})}
\,\rho_{\rm R}(T_{\rm in}),
\end{equation}
where $\mathcal{H}$ is the Hubble expansion rate and $\gamma$ accounts for the efficiency of gravitational collapse. Numerical studies suggest a representative value $\gamma \simeq 0.2$ \cite{Carr:1974nx}. The corresponding formation time and temperature can be obtained as
\begin{eqnarray}
    t_{\rm in} &=& \frac{M_{\rm in}}{8\, \pi \gamma M_{P}^2}, \\ 
    T_\text{in} &=& \Biggl(\frac{1440\,\gamma^2}{g_*\left(T_\text{in}\right)}\Biggr)^{1/4}\,\sqrt{\frac{M_P}{M_\text{in}}}\,M_P\,,
\end{eqnarray}
with $M_{P}$ denoting the reduced Planck mass. The
upper bound on the inflationary scale \cite{Planck:2018jri, BICEPKeck:2021gln} provides a lower bound on the initial PBH mass $M_{\rm in} \gtrsim 0.1 \text{g}$.

Following their formation, PBH lose mass through Hawking evaporation by emitting particles into the surrounding thermal bath \cite{Hawking:1974rv, Hawking:1975vcx}. Here, we use $M_{\rm in}$ to denote the initial mass of the black hole at formation, while $M_{\rm BH}$
denotes the black hole mass at a later stage of its evolution. At a particular instance of evolution, the Hawking temperature and Bekenstein--Hawking entropy possessed by PBH are given by
\begin{eqnarray}
T_{\rm BH}
&=&
\frac{M_P^2}{M_{\rm BH}},
\\[2mm]
S
&=&
\frac{1}{2}
\left(
\frac{M_{\rm BH}}{M_P}
\right)^2
=
\frac{1}{2}
\left(
\frac{M_P}{T_{\rm BH}}
\right)^2 ,
\label{eq:entropy}
\end{eqnarray}
respectively. 

The net mass loss rate of a PBH due to Hawking evaporation can be expressed as \begin{eqnarray} \label{eq:massloss}
    \frac{dM_{\rm BH}}{dt} = \dot{M}_{\rm BH} \simeq - \epsilon \frac{M^4_{P}}{M^2_{\rm BH}},
\end{eqnarray}
where $\epsilon$ carries the information of the emitted particles and is given as
\begin{equation} 
\epsilon =  \frac{ 3.8\, \pi g_{*, H}(T_{\rm BH})}{480}, \,\, g_{*, H}(T_{\rm BH}) = \sum_i \omega_i g_{i,H}, \,\,g_{i,H}=
    \begin{cases}
        1.82
        &\text{for }s=0\,,\\
        1.0
        &\text{for }s=1/2\,,\\
        0.41
        &\text{for }s=1\,,\\
        0.05
        &\text{for }s=2\,,\\
    \end{cases}
    \label{eq:gsh}
\end{equation}
with $\omega_i=2s_i+1$ for massive particles of spin $s_i$, $\omega_i=2$ for massless species with $s_i>0$, and $\omega_i=1$ for spinless species. Integrating Eq. \eqref{eq:massloss} and assuming radiation domination throughout the evolution of a PBH, its corresponding evaporation temperature in terms of initial mass can be calculated as
\begin{eqnarray} \label{eq:Tev}
    T_{\rm ev} = \left(\frac{40}{\pi^2 g_{*}(T_{\rm ev})}\right)^{1/4} \left(\frac{ 3 \, \epsilon\, M^{5}_{\rm P}}{M^{3}_{\rm in}}\right)^{1/2}.
\end{eqnarray}
However, if PBH dominate the energy density of the Universe at some point in their evolution, the bath temperature after their complete evaporation becomes $\bar{T}_{\rm ev} = \frac{2}{\sqrt{3}} T_{\rm ev}$. Assuming PBH domination, an upper bound on PBH mass $M_{\rm in} \lesssim 4\times 10^{8} \, \text{g}$ is obtained by demanding that PBH evaporate before the onset of Big Bang Nucleosynthesis (BBN), $\bar{T}_{\rm ev} \gtrsim T_{\rm BBN} \sim 4 \,\text{MeV}$. Hence, throughout this work, we restrict our analysis to primordial black holes in the mass range: $0.1\, \text{g} \lesssim M_{\rm in} \lesssim 4\times 10^{8} \, \text{g} $ with an initial abundance ($\beta$) large enough for PBH to dominate the energy density of the Universe prior to their evaporation.

\subsection{Non-thermal leptogenesis}
As mentioned earlier, we consider a type-I seesaw mechanism where three gauge singlet RHNs {$N_i, i\in [1-3]$} are added to the SM. The relevant part of the Lagrangian is given by
\begin{equation}
    -\mathcal{L}_{N} \supset (Y_\nu)_{\alpha i} \overline{\ell_{L\alpha}} \tilde{H} N_i + \frac{1}{2} m_{N_i} \overline{N^c_i} N_i + {\rm h.c.}
    \label{eq:10}
\end{equation}
where $\ell_L, H$ denote lepton and Higgs doublet of the SM, respectively. Once the electroweak symmetry gets spontaneously broken after the Higgs acquires a non-zero vacuum expectation value (VEV) $v$, light neutrino masses arise in the seesaw limit $M_D = Y_\nu v/\sqrt{2} \ll m_N$ as 
\begin{equation}
    m_\nu = -M_D m^{-1}_N M^T_D = -Y_\nu m^{-1}_N Y^T_\nu \frac{v^2}{2}.
\end{equation}

In thermal leptogenesis, RHNs are produced from the thermal bath due to the Yukawa interactions. In non-thermal leptogenesis setup we are considering here, RHNs are produced from PBH evaporation. There have been several works in the literature where such role of PBH evaporation  on generating baryon asymmetry of the Universe has been studied \cite{Hawking:1974rv, Carr:1976zz, Baumann:2007yr, Hook:2014mla, Fujita:2014hha, Hamada:2016jnq, Morrison:2018xla, Hooper:2020otu, Perez-Gonzalez:2020vnz, Datta:2020bht, JyotiDas:2021shi, Smyth:2021lkn, Barman:2021ost, Bernal:2022pue, Ambrosone:2021lsx}. Due to gravitational interactions, the evaporation of PBH produces all particles, including the RHNs. Here, we consider a scenario where the RHNs produced from PBH evaporation never reach thermal equilibrium, such that we remain in the non-thermal leptogenesis regime. Subsequently, the CP-violating decay of these RHNs generates a non-zero lepton asymmetry which gets converted into baryon asymmetry via $B+L$-violating electroweak sphaleron transitions \cite{Kuzmin:1985mm}. 

In order to compute the PBH-assisted baryon asymmetry, we need to track the energy densities of PBH ($\rho_{\rm BH}$), radiation ($\rho_{\rm R}$), RHNs ($ \rho_{\rm N}$), as well as the bath temperature ($T$) throughout the PBH evolution. The corresponding Boltzmann Equations (BEs) are 
\begin{eqnarray} \label{eq:BE_PBH}
    \frac{d \rho_{\rm BH}}{dt} &+& 3\,\mathcal{H} \rho_{\rm BH} = \frac{\dot{M}_{\rm BH}}{M_{BH}} \rho_{\rm BH}, \nonumber \\ 
    \frac{d \rho_{\rm R}}{dt} &+& 4\,\mathcal{H} \rho_{\rm R} = - \frac{\epsilon_{\rm SM}}{\epsilon_{\rm SM}+\epsilon_{\rm N}} \frac{\dot{M}_{\rm BH}}{M_{BH}} \rho_{\rm BH}, \nonumber \\
    \frac{d \rho_{\rm N}}{dt} &+& 3\,\mathcal{H} \rho_{\rm N} = - \frac{\epsilon_{\rm N}}{\epsilon_{\rm SM}+\epsilon_{\rm N}} \frac{\dot{M}_{\rm BH}}{M_{BH}} \rho_{\rm BH}, \nonumber \\
    \mathcal{H}^{2} &=& \frac{\rho_{\rm BH} + \rho_{\rm R} + \rho_{\rm N}}{3 M^2_{\rm P}}, \nonumber \\
    \frac{\dot{T}}{T} &=& - \frac{1}{\Delta} \left(\mathcal{H} + \frac{\epsilon_{\rm SM}}{4 (\epsilon_{\rm SM}+\epsilon_{\rm N})} \frac{\rho_{\rm BH}}{\rho_{\rm R}} \frac{\dot{M}}{M} \frac{g_{*}(T)}{g_{*,s}(T)}\right).
\end{eqnarray}
Here $^{.}$ denotes derivatives with respect to cosmic time and  $\Delta = 1 + \frac{T}{3} \frac{d\, g_{*,s}(T)/dT}{g_{*,s}(T)}$. The quantities $\epsilon_{\rm SM}$ and $\epsilon_{\rm N}$ carry the information of PBH emitted particles for SM and RHN as described in Eq. \eqref{eq:gsh}. By solving the BEs given in Eqs. \eqref{eq:BE_PBH}, the number density of $N_{1}$ is obtained at the time of PBH evaporation. Finally, the total baryon asymmetry today from PBH-assisted leptogenesis can be expressed as 
\begin{eqnarray} \label{eq:asymmetry1}
    Y^{\Delta L} ={\varepsilon^{\rm CP}_{1}} a_{\rm sph} \frac{n_{\rm N_{1}}(T_{\rm ev})}{s (T_{\rm ev})},
\end{eqnarray}
where $a_{\rm sph}= 28/79$ is the sphaleron conversion factor. Here, the CP asymmetry parameter is : 
\begin{eqnarray}
    {\varepsilon^{\rm CP}_{1}} = \frac{1}{8\pi} \sum_{j=2,3} \frac{\text{Im}[(Y^{\dagger}_{\nu}Y_{\nu})^2_{1j}]}{(Y^{\dagger}_{\nu}Y_{\nu})_{11}}\left\{{f_{V}\left(\frac{m^2_{N_{j}}}{m^2_{N_{1}}}\right)+f_{S}\left(\frac{m^2_{N_{j}}}{m^2_{N_{1}}}\right)}\right\},
\end{eqnarray}
where $Y_{\nu}$ is the Yukawa matrix corresponding to the term $(Y_{\nu})_{\alpha i} \bar{\ell}_{L\alpha}\tilde{H} N_{i}$ and $f_{V}(x) = \sqrt{x}\left[1-(1+x)\text{ln}\left(\frac{1+x}{x}\right)\right]$, and $f_{S}(x) = \frac{\sqrt{x}}{1-x}$ denote the contributions of vertex and self-energy correction respectively. In the hierarchical RHN mass limit of $m_{N_{1}} \ll m_{N_{2,3}}$, 
\begin{eqnarray}
    f_{V}(x) + f_{S}(x) \simeq - \frac{3}{2\sqrt{x}}, \,\,\,\,\,\,\, \text{for}\,\,\,\,  x\gg 1.
\end{eqnarray}

Moreover, the upper bound on the CP asymmetry parameter can be connected with the low-energy neutrino parameter as
\begin{eqnarray}
    {\varepsilon^{\rm CP}_{1}}| \lesssim \frac{3 m_{N_{1}} \sqrt{(\Delta m_{\rm atm})^2}}{8 \pi v^{2}}.
\end{eqnarray}

\subsubsection{Analytical estimation of Baryon Asymmetry}

 In terms of the number of the lightest RHN $N_{1}$ species, $N_{N_{1}}$, produced from the complete evaporation of a PBH, the Eq. \eqref{eq:asymmetry1} can be expressed as 
\begin{eqnarray}
     Y^{\Delta L} = {\varepsilon^{\rm CP}_{1}} a_{\rm sph} \frac{N_{\rm N_{1}} n_{\rm BH}(a_{\rm ev})}{s (a_{\rm ev})}.
\end{eqnarray}
The expression for $N_{N_{1}}$ is obtained from the generic expression of the number of a species with mass $m_{X}$ and internal degrees of freedom $g_{X}$ produced from the complete evaporation of a PBH, which is given as
\begin{eqnarray} \label{eq:Nx}
    N_{X} \simeq \frac{27}{128} \frac{\xi g_{j} \zeta(3)}{\pi^3 \epsilon}  \begin{cases}
    \left(\frac{M_{\rm in}}{M_P}\right)^{2},& \text{if $m_{X} < T^{\rm in}_{\rm BH}$} \\
    \left(\frac{M_P}{m_{X}}\right)^{2},& \text{if $m_{X} > T^{\rm in}_{\rm BH}$}.
\end{cases}
\end{eqnarray}
Here, $\xi$ takes the value of $1$ for bosons and $3/4$ for fermions. {Using the relation $\rho_{\rm BH}(a_{\rm ev}) = \rho_{\rm R}(a_{\rm ev}) \simeq n_{\rm BH}(a_{\rm ev})\times M_{\rm in}$, the number density of PBH at $a=a_{\rm ev}$ is found to be }
\begin{eqnarray} \label{eq:naev}
    n_{\rm BH} (a_{\rm ev}) = 12\, M^3_{\rm P}\, \epsilon^2 \left(\frac{M_{P}}{M_{\rm in}}\right)^7.
\end{eqnarray}
Combining the expression of Eqs. \eqref{eq:Nx}, \eqref{eq:naev} and \eqref{eq:Tev} in Eq. \eqref{eq:asymmetry1}, the PBH assisted baryon asymmetry can be approximately written as 
\begin{eqnarray} \label{eq:lepto_analy}
    Y^{\Delta L} \simeq 1.04\times10^{-2}\, \frac{g^{3/4}_{*}(T_{\rm ev})}{g_{*,s}(T_{\rm ev})}\, {\varepsilon^{\rm CP}_{1}}\, a_{\rm sph}\, \begin{dcases}
        \left(\frac{M_{\rm P}}{M_{\rm in}}\right)^{1/2} ,& \text{if $m_{N_{1}} < \,{T^{\rm in}_{\rm BH}}$} \\
        \left(\frac{M^{9}_{\rm P}}{M^{5}_{\rm in} m^{4}_{N_{1}}}\right)^{1/2} ,& \text{if $m_{N_{1}} > \,T^{\rm in}_{\rm BH}$}.
    \end{dcases}
\end{eqnarray}

\begin{figure}
    \centering
    \includegraphics[width=0.49\linewidth]{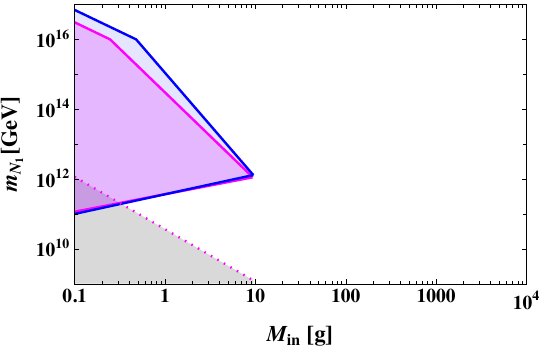}
    \includegraphics[width=0.49\linewidth]{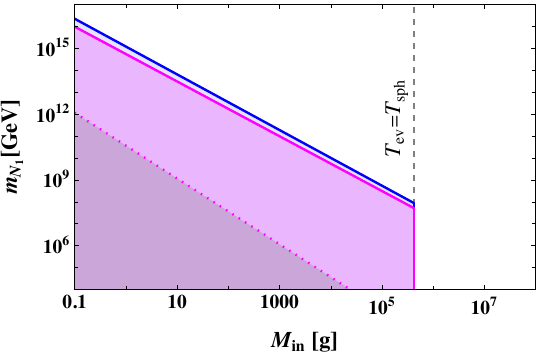}
    \caption{Allowed regions in the $M_{\rm in}$ -- $m_{N_1}$ plane for hierarchical leptogenesis (left panel) and resonant leptogenesis (right panel). {The magenta-shaded areas correspond to parameter points yielding successful leptogenesis using analytical approximations, while the regions enclosed by the blue lines are obtained from numerical solutions.} The bottom left triangular gray shaded regions correspond to $m_{N_{1}}<T_{\rm ev}$, inconsistent with the non-thermal leptogenesis requirement.}
    \label{fig:analytical_lepto}
\end{figure}


{The left panel of Fig. \ref{fig:analytical_lepto} shows the viable parameter space for a hierarchical RHN mass spectrum in the $M_{\rm in}$ -- $m_{N_{1}}$ plane, obtained using analytical approximations (regions enclosed by magenta curves) and numerical solutions (regions enclosed by blue curves).} The observed analytical behavior can be understood from Eq. \eqref{eq:lepto_analy}. In the regime $m_{N_{1}} < T^{\rm in}_{\rm BH}$, the generated baryon asymmetry scales as $Y^{\Delta L} \propto \frac{m_{N_{1}}}{\sqrt{M_{\rm in}}},$ implying a positive correlation between $M_{\rm in}$ and $m_{N_{1}}$ along the lower boundary of the allowed region. In contrast, for $m_{N_{1}} > T^{\rm in}_{\rm BH}$, the asymmetry scales as $Y^{\Delta L} \propto \frac{1}{M^{5/2}_{\rm in} m_{N_{1}}},$ leading to an anti-correlation between $M_{\rm in}$ and $m_{N_{1}}$ along the upper boundary. These scaling behaviors explain the distinct slopes of the lower and upper limits of the viable parameter space. We find that the allowed PBH mass range in the hierarchical scenario is approximately $0.1~{\rm g} \lesssim M_{\rm in} \lesssim 10~{\rm g}$. {The discrepancy between the analytical and numerical solutions arises for $m_{N_{1}} > T^{\rm in}_{\rm BH}$, as the analytical approximation does not adequately capture RHN production when $T_{\rm BH} \sim m_{N_{1}}$.}  Additionally, the bottom gray-shaded region denotes the parameter space incompatible with non-thermal leptogenesis, obtained by approximately imposing the condition $m_{N_{1}}< T_{\rm ev}$. 

The generated lepton asymmetry can be substantially enhanced when two RHNs are nearly degenerate in mass, i.e., $\Delta m \equiv m_{N_2}-m_{N_1} \ll \bar{m} \equiv (m_{N_1}+m_{N_2})/2$, giving rise to the resonant leptogenesis scenario \cite{Pilaftsis:2003gt}. In particular, when the mass splitting becomes comparable to the RHN decay width, $\Delta m \simeq \Gamma$, the CP asymmetry parameter can be resonantly amplified and reach values of $\mathcal{O}(1)$. As a consequence, successful leptogenesis can be achieved over a significantly larger region of parameter space, as illustrated in the right panel of Fig. \ref{fig:analytical_lepto} where we consider the CP asymmetry to be $0.1$. The vertical dashed lines indicate the condition $T_{\rm ev}=T_{\rm sph}$, corresponding to the epoch of sphaleron decoupling.

\section{Gravitational wave signatures}
\label{sec2}

\subsection{Gravitational Bremsstrahlung}
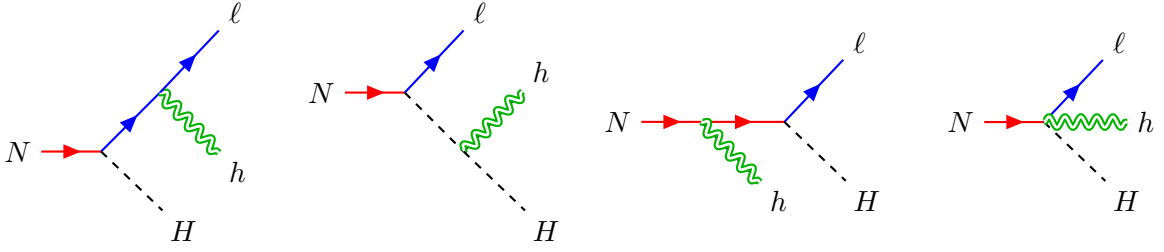
\begin{figure}[t]
	\centering
	\begin{subfigure}[c]{.24\textwidth}
		\begin{tikzpicture}
			\begin{feynman}
				\vertex (a) {\(N\)};
				\vertex [right=1.1cm of a] (b) ;
				\vertex [above right=1.1cm of b] (c);
				\vertex [above right=1.1cm of c] (d) {\(\ell\)};
				\vertex [below right=1.1cm of c] (e) {\(h \)};
				\vertex [below right=1.1cm of b] (f) {\(H \)};
				\diagram* { 
                (a) -- [fermion,thick,red] (b),
                (b) -- [fermion,thick,blue] (c),
                (c) -- [  thick, green!70!black, double distance=1pt, decoration={snake,amplitude=2pt,segment length=6pt}, decorate] (e),
                (c) -- [fermion,thick,blue] (d),
                (b) -- [dashed,thick,black] (f)
                };
				\end{feynman}
		\end{tikzpicture}
	\end{subfigure}	
	\begin{subfigure}[c]{.24\textwidth}
		\begin{tikzpicture}
			\begin{feynman}
				\vertex (a) {\(N\)};
				\vertex [right=1.1cm of a] (b) ;
				\vertex [above right=1.1cm of b] (c) {\(\ell\)};
				\vertex [below right=1.1cm of b] (d) ;
				\vertex [below right=1.1cm of d] (f) {\(H \)};
				\vertex [above right=1.1cm of d] (e) {\(h \)};
				\diagram* {
    (a) -- [fermion, thick, red] (b),

    (b) -- [fermion, thick, blue] (c),

    (b) -- [dashed, thick, black] (d),

    (d) -- [dashed, thick, black] (e),

    (d) -- [
        thick,
        green!70!black,
        double distance=1pt,
        decoration={snake, amplitude=2pt, segment length=6pt},
        decorate
    ] (e),

    (d) -- [dashed, thick, black] (f)
};
				\end{feynman}
		\end{tikzpicture}
	\end{subfigure}
	\begin{subfigure}[c]{.24\textwidth}
	\begin{tikzpicture}
		\begin{feynman}
			\vertex (a) {\(N\)};
			\vertex [right=1.1cm of a] (b) ;
			\vertex [right=1.1cm of b] (d) ;
			\vertex [below right=1.1cm of b] (c) {\(h\)};
			\vertex [above right=1.1cm of d] (e) {\(\ell\)};
			\vertex [below right=1.1cm of d] (f) {\(H\)};
			\diagram* {
    (a) -- [fermion, thick, red] (b),

    (b) -- [fermion, thick, red] (d),

    (b) -- [dashed, thick, black] (c),

    (b) -- [
        thick,
        green!70!black,
        double distance=1pt,
        decoration={snake, amplitude=2pt, segment length=6pt},
        decorate
    ] (c),

    (d) -- [fermion, thick, blue] (e),

    (d) -- [dashed, thick, black] (f)
};
			\end{feynman}
	\end{tikzpicture}
\end{subfigure}
	\begin{subfigure}[c]{.24\textwidth}
		\begin{tikzpicture}
			\begin{feynman}
				\vertex (a) {\(N\)};
				\vertex [right=1.1cm of a] (b) ;
				\vertex [above right=1.1cm of b] (c){\(\ell \)};
				\vertex [right=1.1cm of b] (d) {\(h \)};
				\vertex [below right=1.1cm of b] (e) {\(H \)};
				\diagram* {
    (a) -- [fermion, thick, red] (b),

    (b) -- [fermion, thick, blue] (c),

    (b) -- [dashed, thick, black] (e),

    (b) -- [
        thick,
        green!70!black,
        double distance=1pt,
        decoration={snake, amplitude=2pt, segment length=6pt},
        decorate
    ] (d)
};
				\end{feynman}
		\end{tikzpicture}
	\end{subfigure}	
	\caption{Feynman diagrams for graviton bremsstrahlung induced by RHN decay. }
	\label{fig:Feyn_diafram}
\end{figure}

As mentioned earlier, graviton bremsstrahlung from the decay of RHN \cite{Datta:2024tne, Choi:2025hqt, Murayama:2025thw, Kanemura:2025rct} offers an interesting GW probe directly connected to the processes responsible for generating lepton asymmetry. The interaction between the SM and BSM sectors is described by the action
\begin{eqnarray}
    S \supset \int d^{4}x \sqrt{-g}
    \left[
    \frac{M_{\rm P}^{2}}{2} R
    + \mathcal{L}_{\rm SM}
    + \mathcal{L}_{N}
    \right],
\end{eqnarray}
where $R$ denotes the Ricci scalar and $g \equiv \det(g_{\mu\nu})$ is the determinant of the spacetime metric. In the weak-field limit, the metric can be expanded around the flat Minkowski background as
\begin{eqnarray}
    g_{\mu\nu} \simeq \eta_{\mu\nu}
    + \frac{2}{M_{\rm P}} h_{\mu\nu},
\end{eqnarray}
where $h_{\mu\nu}$ represents the canonically normalized graviton field. The corresponding interaction between the graviton and the energy-momentum tensor $T_i^{\mu\nu}$ of a generic particle species $i$ is given by
\begin{eqnarray}
    \mathcal{L}_{\rm int}^{\rm gravity}
    \supset
    -\frac{2}{M_{\rm P}}
    h_{\mu\nu} T_i^{\mu\nu}.
\end{eqnarray}
The leading-order processes contributing to the GW spectrum through gravitational bremsstrahlung accompanying the decay of RHNs are depicted in Fig.~\ref{fig:Feyn_diafram}. As the graviton is emitted from one of the external legs of the same RHN decay process relevant for leptogenesis, this offers a powerful and unavoidable probe of leptogenesis.

 In order to obtain the  spectrum of GW produced via bremsstrahlung, we first need to solve the BE for the energy density of graviton, $\rho_{\rm gw}$, which is given as \cite{Murayama:2025thw, Datta:2024tne}

\begin{eqnarray} 
    \frac{d \rho_{\rm gw}}{dt}  + 4 \mathcal{H} \rho_{\rm gw} = \int\frac{d^3 k} {(2\pi)^{3}} E_{k} \frac{1}{2 E_{k}} \int\frac{d^3 p}{(2\pi)^{3} \,2 E_{p}} \int\frac{d^3 q}{(2\pi)^{3}\, 2 E_{q}} \int\frac{d^3 r}{(2\pi)^{3} \, 2E_{r}} |\mathcal{M}|^{2} f_{N}(p) \nonumber \\ \times (2\pi)^4 \delta^{(4)}(P-Q-K-R),
\end{eqnarray}
 where $P,\, Q,\, R,\, K$ denote the four momentum of RHN, lepton, Higgs, and graviton, respectively. $f_{N}$ is the distribution function of RHN and $|\mathcal{M}|^{2}$ represents the spin and polarization averaged matrix element squared. The Eq. \eqref{eq:BE_gw} can be written as  
\begin{eqnarray} 
    \frac{d \rho_{\rm gw}}{dt}  + 4 \mathcal{H} \rho_{\rm gw} &=& \int\frac{d^3 p}{(2\pi)^{3}} f_{N}(p)\, \frac{m_{N_{1}}}{E_{N_{1}}}\int \frac{d\Gamma^{1\to3}}{dE_{k}} E_{k}\, dE_{k}
\end{eqnarray}
which in the non-relativistic limit can be written as 
\begin{eqnarray} \label{eq:BE_gw}
    \frac{d \rho_{\rm gw}}{dt}  + 4 \mathcal{H} \rho_{\rm gw} &=& \left[\int \frac{d \Gamma ^{1 \to 3}}{d E_{\rm k}}\left(\frac{E_{\rm k}}{m_{\rm N_{1}}}\right)dE_{\rm k}\right] \rho_{N_{1}}.
\end{eqnarray}
Here,
\begin{eqnarray} \label{eq:3_decay_width}
    \frac{d\Gamma^{1\to 3}}{dE_{\rm k}}= \frac{(Y_{\nu}^{\dagger}Y_{\nu})_{11}}{256 \pi^{3}} \frac{m^2_{N_{1}}}{M^2_{P}} \frac{(x-1)^2(1-2x)}{x},
\end{eqnarray}
with $x = E_{k}/m_{N_{1}}$. 

Calculating $\rho_{\rm gw}$ requires solving Eqs. \eqref{eq:BE_gw} and \eqref{eq:3_decay_width} along with the coupled BEs in Eqs. \eqref{eq:BE_PBH}. We solve these equations numerically. Below, we also provide an analytical estimation of the calculation of $\Omega_{\rm GW}h^2$. 

\subsubsection{Analytical estimation of GW  via bremsstrahlung}

The Eq. \eqref{eq:BE_gw} can be re-written as 
\begin{eqnarray}
    \frac{d}{dt} \left(\frac{d\rho_{\rm gw}}{d E_{k}}\right) + 4 \mathcal{H} \frac{d\rho_{\rm gw}}{d E_{k}} = \frac{d \Gamma ^{1 \to 3}}{d E_{\rm k}} E_{\rm k}\, n_{N_{1}}.
\end{eqnarray}
In terms of scale factor $a$,
\begin{eqnarray}
    \frac{d}{da} \left(a^5 \frac{d\rho_{\rm gw}}{d \, \text{ln} \, E_{k}}\right) = \frac{a^4}{\mathcal{H}} \frac{d \Gamma ^{1 \to 3}}{d E_{\rm k}} E^2_{\rm k}\, n_{N_{1}}.
\end{eqnarray}
Integrating the above equation from the evaporation of PBH, $a_{\rm ev}$ to the decay of RHN, $a_{N}$, we get
\begin{eqnarray}
    \frac{d\rho_{\rm gw}(a_{\rm N})}{d \, \text{ln}\, E_{k}} \simeq \frac{2}{3} \tau_{N}\, n_{N_{1}}(a_{\rm ev}) \left(\frac{a_{\rm ev}}{a_{N}}\right)^3 \left[1-\left(\frac{a_{\rm ev}}{a_{\rm N}}\right)^3\right]\left.\left(\frac{d \Gamma ^{1 \to 3}}{d E_{\rm k}}\right)\right|_{a_{N}} E^2_{\rm k} (a_{\rm N}),
\end{eqnarray}
where $\tau_{N}$ is the decay lifetime of RHN. Hence, the GW amplitude today can be written as 
\begin{eqnarray}
    \Omega_{\rm GW}h^2 &=& \frac{h^2}{\rho_{\rm c, 0}} \frac{d\rho_{\rm gw}(a_{\rm N})}{d \, \text{ln}\, E_{k}} \left(\frac{a_{\rm N}}{a_{0}}\right)^{4} \\ \nonumber
    &\simeq& \frac{h^2}{\rho_{\rm c, 0}} \frac{2}{3} \tau_{N}\, n_{N_{1}}(a_{\rm ev}) \left.\left(\frac{d \Gamma ^{1 \to 3}}{d E_{\rm k}}\right)\right|_{a_{N}} E^2_{\rm k} (a_{\rm N}) \left(\frac{a_{\rm ev}}{a_{N}}\right)^3 \left[1-\left(\frac{a_{\rm ev}}{a_{\rm N}}\right)^3\right] \left(\frac{a_{\rm N}}{a_{0}}\right)^{4},
\end{eqnarray}
with, $\rho_{c,0}$ denoting the critical energy density today. The frequency today, $f$, is related to the energy of the graviton at the time of RHN decay as
\begin{eqnarray}
    E_{k}(a_{\rm N}) = 2 \pi f \left(\frac{a_{0}}{a_{\rm N}}\right).
\end{eqnarray}
Simplifying the expression for $\Omega_{\rm GW}h^2$ in terms of $f$ for $m_{N_{1}}<T^{\rm in}_{\rm BH}$, we obtain
\begin{eqnarray} \label{eq:GWs_amp}
    \Omega_{\rm GW}h^2 \simeq 2.7\times10^{-20} \left(\frac{f}{10^{10} \, \text{Hz}}\right) \left(\frac{m_{N_{1}}}{10^{15}\, \text{GeV}}\right)^2 \sqrt{\frac{1\, \text{g}}{M_{\rm in}}}.
\end{eqnarray}
For the scenario where $m_{N_{1}}>T^{\rm in}_{\rm BH}$, the amplitude takes the form of
\begin{eqnarray} \label{eq:GWs_amp_1}
    \Omega_{\rm GW}h^2 \simeq 3\times 10^{-24} \left(\frac{f}{10^{10}\, \text{Hz}}\right) \left(\frac{1 \, \text{g}}{M_{\rm in}}\right)^{5/2}.
\end{eqnarray}
Moreover, as the graviton can carry the maximum energy of $m_{N_{1}}/2$ at the time of formation, the peak GW frequency can be written as 
\begin{eqnarray}
    f^{\rm peak} &=& \frac{m_{N_{1}}}{4\pi} \left(\frac{a_{N_{1}}}{a_{0}}\right) \\ \nonumber &=& \frac{m_{N_{1}}}{4\pi} \left(\frac{T^3_0\, g_{*,s}(T_{0})}{T^3_N\, g_{*,s}(T_{N})}\right)^{1/3},                  
\end{eqnarray}
where $T_{N}$ denotes the temperature when $N_{1}$ decays. Using the relation $\mathcal{H}=\frac{1}{2 \tau_{N}}$, $T_{N}$ can be determined as
\begin{eqnarray}
    T_{N} = \left(\sqrt{\frac{90}{g_{*,s}(T_{N})}}\,\frac{(Y^{\dagger}_{\nu} Y_{\nu})_{11} m_{N_{1}} M_{\rm P}}{16\pi^2} \right)^{1/2}.
\end{eqnarray}
Consequently, the peak frequency can be expressed as 
\begin{eqnarray} \label{eq:peak_f}
    f^{\rm peak} = 1.2\times 10^{16}\, \text{Hz} \left(\frac{10^{-5}}{(Y^{\dagger}_{\nu} Y_{\nu})^{1/2}_{11}}\right) \left(\frac{m_{N_{1}}}{M_{P}}\right)^{\frac{1}{2}}.
\end{eqnarray}

\begin{figure}
    \centering
    \includegraphics[width=0.65\linewidth]{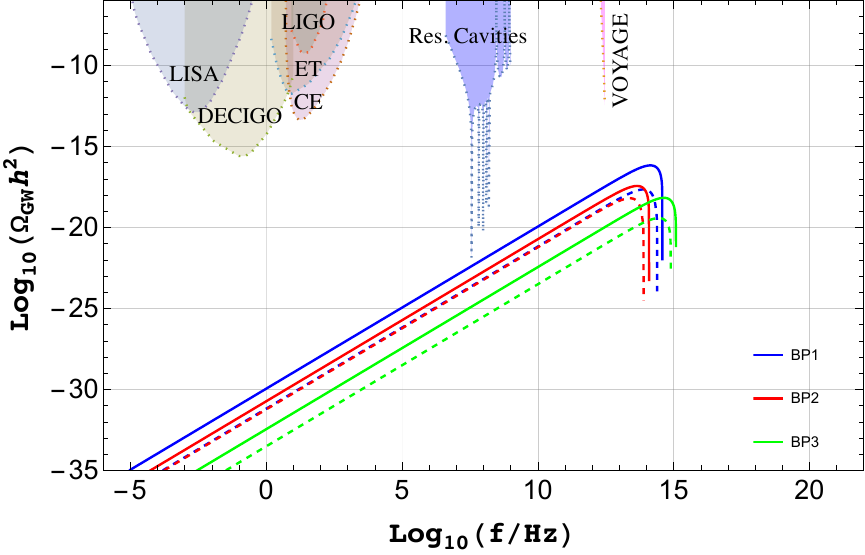}
    \refstepcounter{footnote}
    \setcounter{figfn}{\value{footnote}}
    \caption{GW spectra via RHN bremsstrahlung corresponding to the benchmark points BP1 (blue), BP2 (red), and BP3 (green). The numerical results are shown by the solid curves, while the dashed curves represent the corresponding analytical approximations. The shaded regions denote the current constraints and projected sensitivities of present and future GW observatories: LIGO~\cite{LIGOScientific:2014pky}, ET~\cite{Punturo_2010}, CE~\cite{LIGOScientific:2016wof}, DECIGO~\cite{Yagi:2011wg}, LISA~\cite{LISACosmologyWorkingGroup:2022kbp}, resonant cavities~\cite{Herman:2022fau}, and VOYAGE~\cite{He:2023xoh}. \protect\footnotemark[\value{figfn}]}
    \label{fig:GW_benchmark}
\end{figure}

\makeatletter
\renewcommand{\thefootnote}{\alph{footnote}}
\makeatother
\footnotetext[\value{figfn}]{
While several proposals for detecting high-frequency GWs have been put forward~\cite{Berlin:2021txa, Berlin:2023grv, Aggarwal:2020umq, Carney:2024zzk}, their projected sensitivities may already saturate the current constraints imposed on $N_{\rm eff}$ from CMB and BBN observations.
}

\begin{table}[htb!]
\centering
\renewcommand{\arraystretch}{1.2}
\begin{tabular}{|c|c|c|c|c|c|c|c|}
\hline
\text{Benchmark} & ${M_{\rm in} \, \text{(g)}}$  & ${\beta}$ &
${m_{N_1}} \, \text{(GeV)}$ & ${m_{N_2}} \, \text{(GeV)}$ &
${m_{N_3}} \, \text{(GeV)}$ & ${\sqrt{(Y^{\dagger}_{\nu}Y_{\nu})_{11}}}$& {$\varepsilon^{\rm CP}_{1}$} \\
\hline
BP1 &  0.1 & $5\times10^{-4}$ & $10^{15}$ & $7\times10^{15}$  & $10^{16}$ &  $10^{-5}$ & {$10^{-3}$}  \\
\hline
BP2 &  0.1 & $5\times10^{-4}$ & $10^{14}$ & $7\times10^{15}$  & $10^{16}$ &  $10^{-5}$ &{ $10^{-5}$} \\
\hline
BP3 & 1 & $5\times10^{-4}$ & $10^{14}$ & $7\times10^{15}$  & $10^{16}$ &  $10^{-6}$& {$3\times 10^{-3}$}  \\
\hline
\end{tabular}
\caption{Benchmark points for hierarchical leptogenesis. }
\label{tab:BP}
\end{table}

Figure~\ref{fig:GW_benchmark} displays the GW spectra generated through bremsstrahlung processes for the three benchmark points (BPs) listed in Table~\ref{tab:BP}. The solid curves correspond to the full numerical results, while the dashed curves represent the analytical approximations. As anticipated from the analytical estimates, the peak frequency shifts toward higher values with increasing $m_{N_1}$ and decreasing Yukawa coupling. A comparison between BP1 and BP2 reveals a small difference. While the analytical approximation predicts nearly identical GW amplitudes at a fixed frequency for these two benchmark points, the numerical results exhibit a slight suppression of the amplitude for BP2. This discrepancy originates from the different evaluations of $n_{N_1}(a_{\rm ev})$ in the analytical and numerical treatments. In particular, the analytical approximation becomes less reliable in the regime where $T_{\rm BH}^{\rm in}\sim m_{N_1}$, resulting in a modest deviation from the full numerical calculation. In contrast, the GW amplitude is noticeably smaller for BP3 than for BP2 at a given frequency. This suppression is primarily due to the larger initial PBH mass, $M_{\rm in}$, associated with BP3. In Fig.~\ref{fig:GW_benchmark}, we restrict our analysis to PBH masses up to $M_{\rm in}=1~{\rm g}$. For heavier PBH, the GW spectrum at a fixed frequency would be further suppressed, as suggested by Eqs.~\eqref{eq:GWs_amp} and \eqref{eq:GWs_amp_1}. Consequently, such spectra are expected to lie below the curves shown in the figure and are therefore not displayed. We stick to very low mass limits of ultra-light PBH such that the graviton bremsstrahlung spectrum remains in the most optimistic range for any possible future detections.

The values of the CP asymmetry parameter required to achieve successful leptogenesis for the benchmark points listed in Table~\ref{tab:BP} are ${\varepsilon^{\rm CP}_{1}}=10^{-3}$, ${\varepsilon^{\rm CP}_{1}}=10^{-5}$, and ${\varepsilon^{\rm CP}_{1}}=3\times10^{-3}$, respectively. In Appendix~\ref{App:A}, we present the Yukawa coupling structures corresponding to specific choices of the lightest neutrino mass and the three complex rotation angles of the orthogonal matrix in Casas--Ibarra parametrisation \cite{Casas:2001sr}. These choices are consistent with the observed neutrino mass spectrum and mixing parameters, while simultaneously reproducing the required values of ${\varepsilon^{\rm CP}_{1}}$ and $(Y_\nu^\dagger Y_\nu)_{11}$ for each benchmark point.

For the resonant leptogenesis scenario, the GW spectrum exhibits the same qualitative features as in the non-resonant case. However, its amplitude is enhanced by approximately a factor of two due to the presence of two nearly degenerate right-handed neutrinos, both of which contribute to the GW production.

\begin{figure}
    \centering
    \includegraphics[width=0.49\linewidth]{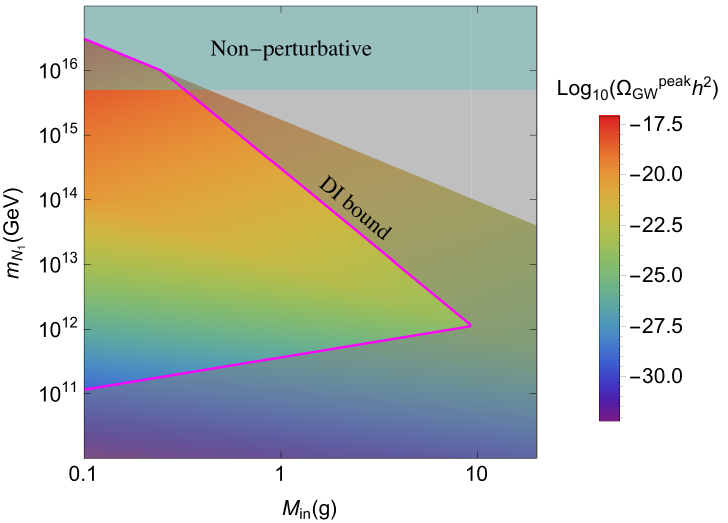}
    \includegraphics[width=0.49\linewidth]{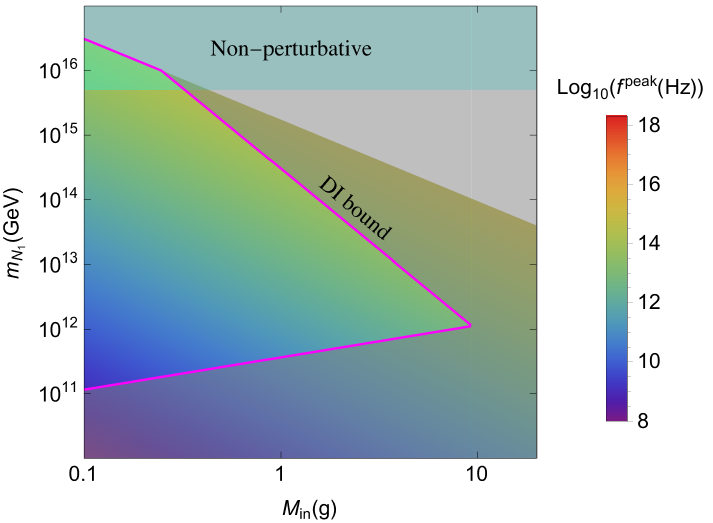}
    \caption{Parameter-space plots for hierarchical leptogenesis in the $M_{\rm in}$ -- $m_{N_1}$ plane. The color bar represents the GW peak amplitude (left panel) and the peak frequency (right panel) considering graviton bremsstrahlung from RHN decay. The parameter space enclosed by the magenta contours yields successful hierarchical leptogenesis, while the gray-shaded region outside the magenta contours is incompatible with the DI bound. The perturbative bound is shown by the horizontal cyan-shaded region.}
    \label{fig:scan1}
\end{figure}

\begin{figure}
    \centering
    \includegraphics[width=0.49\linewidth]{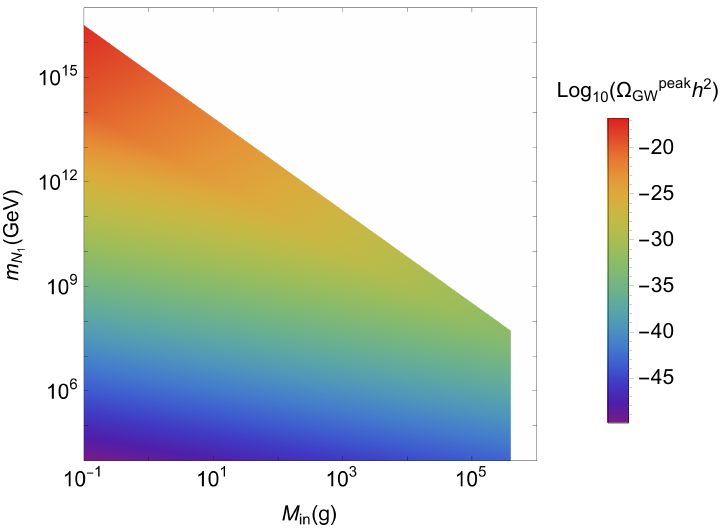}
    \includegraphics[width=0.49\linewidth]{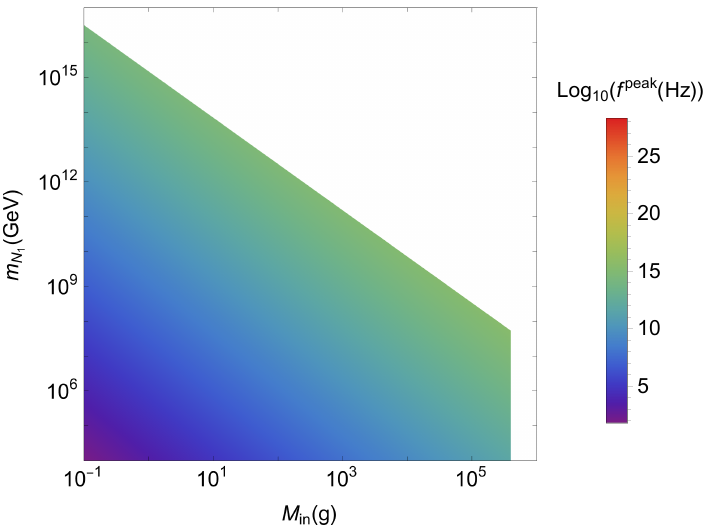}
    \caption{Parameter-space plots for resonant leptogenesis in the $M_{\rm in}$ -- $m_{N_1}$ plane. The color bar represents the GW peak amplitude (left panel) and the peak frequency (right panel) considering graviton bremsstrahlung from RHN decay.}
    \label{fig:scan2}
\end{figure}

In Fig. \ref{fig:scan1}, we present the peak GW amplitude, $\Omega_{\rm GW}^{\rm peak}h^2$, and the corresponding peak frequency, $f^{\rm peak}$, in the $M_{\rm in}$--$m_{N_1}$ plane. Throughout this analysis, we fix the lifetime of $N_1$ such that its decay lifetime is ten times the PBH lifetime. We also assume a hierarchical RHN mass spectrum with $m_{N_2}=10\,m_{N_1}$ and $m_{N_3}=50\,m_{N_1}$. The parameter space enclosed by the magenta contours yields successful leptogenesis with hierarchical RHNs, while the gray-shaded region outside the magenta contours is incompatible with the DI bound. Imposing the perturbativity condition
$\max \!\left[(Y_{\nu}^{\dagger}Y_{\nu})_{ii}\right] < 4\pi$,
we obtain the corresponding upper bound on $m_{N_{1}}$, shown by the horizontal cyan-shaded region. Analogous results for resonant leptogenesis are presented in Fig. \ref{fig:scan2} where we consider $m_{N_{2}}\approx m_{N_{1}}$ and $m_{N_{3}} = 50\, m_{N_{1}}$.  

From Figs. \ref{fig:scan1} and \ref{fig:scan2}, it is evident that, over most of the parameter space, the predicted GW peak frequency lies beyond the sensitivity reach of both current and planned GW observatories. Furthermore, even in the regions where the peak frequency falls within the projected sensitivity range of future detectors, the corresponding GW amplitude is too small to produce an observable signal.

\subsection{GW from thermal bath}

Apart from the production of GW from the decay of heavy RHNs, a stochastic GW background can also be generated through bremsstrahlung processes in the thermal plasma. This background is commonly referred to as the Cosmic Gravitational Microwave Background (CGMB) \cite{Ghiglieri:2015nfa,Ghiglieri:2020mhm,Ringwald:2020ist,Murayama:2025thw}. The production of CGMB starts at very high temperatures, where graviton emission from the thermal bath is efficient. The evolution of the graviton energy density is governed by the Boltzmann equation
\begin{eqnarray}
    \frac{d \rho_{\rm gw}}{dt}  + 4 \mathcal{H} \rho_{\rm gw} &=& \int \frac{d^3k}{(2\pi)^3} \frac{4}{M^2_{\rm P}} T^4 \,\hat{\eta}\left(T,\frac{k}{T}\right),
\end{eqnarray}
where $\hat{\eta}$ denotes the dimensionless source function defined in Ref.~\cite{Ringwald:2020ist}. 

Expressing the above equation in terms of the entropy density $s$ and neglecting the variation of $g_{*,s}$ at temperatures well above the electroweak scale, one obtains
\begin{eqnarray}
    \frac{d}{dt} \left(\frac{\rho_{\rm gw}}{s^{4/3}}\right) = \frac{4}{2\pi^2 M^2_{\rm P}} \frac{1}{s^{4/3}} \int^{\infty}_{0} k^2 dk \, T^4\,  \hat{\eta}\left(T,\frac{k}{T}\right).
\end{eqnarray}
The dominant production of thermal gravitons takes place between the end of reheating and the onset of PBH domination. Integrating over this period yields
\begin{eqnarray}
    \frac{\rho_{\rm gw}(T_{\rm E})}{s^{4/3} (T_{\rm E})} = \frac{12 \sqrt{10}}{2\pi^3 M_{\rm P}} \frac{1}{\left(\frac{2\pi^2}{45}\right)^{4/3}} \int^{T_{\rm RH}}_{T_{\rm E}} \frac{dT}{T^3 g^{4/3}_{*,s}(T) g^{1/2}_{*,s}}  \int^{\infty}_{0} k^2 dk \,   \hat{\eta}\left(T,\frac{k}{T}\right),
\end{eqnarray}
where $T_{\rm RH}$ and $T_{\rm E}$ denote the reheating temperature and the temperature at the onset of PBH domination, respectively. In terms of per logarithimic wavenumber interval and using the fact that the $\hat{\eta}$ is weakly dependent on temperature,  the above equation becomes
\begin{eqnarray}
    \frac{d \rho_{\rm gw}}{d\, \text{ln}\, k_{\rm E}} (T_{\rm E}, k_{\rm E}) &=&  \frac{12 \sqrt{10}}{2\pi^3 M_{\rm P}} g^{1/3}_{*,s} (T_{\rm E}) T^4_{\rm E} \frac{k^3_{\rm E}}{T^3_{\rm E}}\int^{T_{\rm RH}}_{T_{\rm E}} \frac{dT}{g^{4/3}_{*,s}(T) g^{1/2}_{*,s}(T)}    \hat{\eta}\left(T,\frac{k_{\rm E}}{T_{\rm E}} \left(\frac{g_{*,s}(T)}{g_{*,s}(T_{\rm E})}\right)^{1/3} \right)\nonumber \\
    &\simeq& \frac{12 \sqrt{10}}{2\pi^3 M_{\rm P}}  T^4_{\rm E} \frac{k^3_{\rm E}}{T^3_{\rm E}}\,  \frac{T_{\rm RH}}{ g^{1/2}_{*,s}(T_{\rm RH})}    \hat{\eta}\left(T_{\rm RH},\frac{k_{\rm E}}{T_{\rm E}} \left(\frac{g_{*,s}(T)}{g_{*,s}(T_{\rm E})}\right)^{1/3} \right).
\end{eqnarray}
Redshifting this contribution to the present epoch gives
\begin{eqnarray}
    \Omega^{\rm CGMB}_{\rm GW}h^2 (k_{\rm E}) \simeq  \frac{h^2}{\rho_{\rm c, 0}}  \left(\frac{T_{0}}{T_{\rm E}}\right)^4 \left(\frac{g_{*,s}(T_{0})}{g_{*,s}(T_{\rm E})}\right)^{4/3} \frac{1}{\alpha^{4/3}} \frac{d\rho_{\rm gw}}{d\, \text{ln}\, k_{\rm E}} (T_{\rm E}, k_{\rm E}).
\end{eqnarray}
Here we use the quantity $\alpha$ to take into account the dilution of CGMB due to PBH domination era. In terms of frequency today, $f= \frac{k_{\rm E}}{2\pi} \left(\frac{T_{0}}{T_{\rm E}}\right) \left(\frac{g_{*,s}(T_{0})}{g_{*,s}(T_{\rm E})}\right)^{1/3} \frac{1}{\alpha^{1/3}}$, the GW spectrum today can be expressed as
\begin{eqnarray}
    \Omega^{\rm CGMB}_{\rm GW}h^2 (f) &\simeq&  \frac{h^2\, T_{0}}{\rho_{\rm c, 0}}  \left(\frac{g_{*,s}(T_{0})}{g_{*,s}(T_{\rm E})}\right)^{\frac{1}{3}} \frac{1}{\alpha^{1/3}} \frac{48 \sqrt{10}}{g^{1/2}_{*}(T_{\rm RH})} \frac{T_{\rm RH}}{M_{\rm P}} f^3\, \hat{\eta} \left(T_{\rm RH},\, \frac{2\pi f}{T_{0} } \left(\frac{g_{*,s}(T_{\rm E})}{g_{*,s}(T_{0})}\right)^{\frac{1}{3}} \alpha^{\frac{1}{3}}\right) \nonumber \\
    &\simeq& 4.69\times10^{-12} \,\frac{1}{\alpha^{1/3}} \, \left(\frac{f}{10^{9} \, \text{Hz}}\right)^3 \left(\frac{T_{\rm RH}}{M_{\rm P}}\right) \hat{\eta} \left(T_{\rm RH},\, 0.056\times\left(\frac{f}{10^{9} \, \text{Hz}}\right) \alpha^{\frac{1}{3}}\right).\nonumber \\
\end{eqnarray}

The entropy dilution factor arising from PBH evaporation is given by \cite{Bernal:2022pue}
\begin{eqnarray}
    \alpha \equiv \frac{s(\tilde{T})}{s(T_{\rm ev})}  = \left(1+ \beta \frac{T_{\rm in}}{T_{\rm ev}}\right)^{3/4},
\end{eqnarray}
where $\tilde{T}$ is the temperature of the SM plasma after PBH evaporation occurs. 

\begin{figure}[h!]
    \centering
    \includegraphics[width=0.65\linewidth]{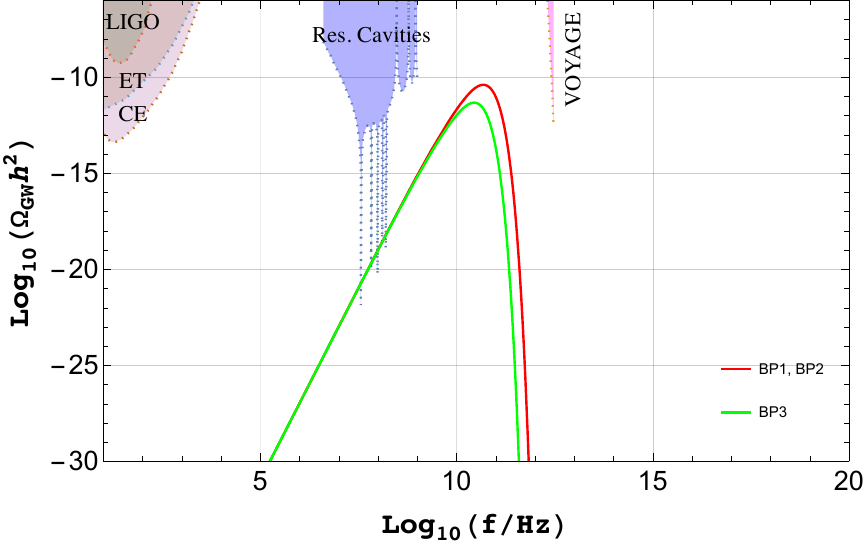}
    \caption{CGMB spectra corresponding to the benchmark points BP1, BP2, and BP3 assuming the maximum allowed reheating temperature, $T_{\rm RH}=10^{16}$ GeV. Since BP1 and BP2 share the same PBH parameters, they yield identical spectra, shown by the red curve. The spectrum corresponding to BP3 is displayed in green.}
    \label{fig:GW_CGMB}
\end{figure}

Fig. \ref{fig:GW_CGMB} shows the CGMB for the benchmark points shown in Table \ref{tab:BP}. As the production of the lightest PBH during the radiation-dominated era requires $T_{\rm RH} \gtrsim 4\times10^{15}\,\mathrm{GeV}$, we take $T_{\rm RH}=10^{16}\,\mathrm{GeV}$, corresponding to the maximum allowed reheating temperature. Comparing these results with those in Fig. \ref{fig:GW_benchmark}, we find that the CGMB spectrum exhibits a lower peak frequency but a larger peak amplitude compared to the ones generated from RHN decay. 

\subsection{Gravitational Waves from PBH}

PBH can generate a stochastic GW background through a variety of physical mechanisms. These include: (i) the emission of gravitons via Hawking evaporation, producing GWs at extremely high frequencies \cite{Anantua:2008am}, (ii)  GWs sourced by density fluctuations due to inhomogeneous distribution of PBH after their formation \cite{Inomata:2019ivs, Papanikolaou:2022chm, Domenech:2021ztg, Domenech:2020ssp, Gouttenoire:2026pmp, Gouttenoire:2026pem}, and (iii) second-order GWs induced by the large primordial scalar perturbations responsible for PBH formation in the early Universe \cite{Saito:2008jc}.  In this work, we focus on the GW backgrounds arising from Hawking evaporation and from density fluctuations associated with the PBH population.

\subsubsection{From density fluctuations}

The stochastic GW background generated by Poisson fluctuations in the PBH number density can be approximated by \cite{Domenech:2020ssp}
\begin{equation}
\Omega_{\rm GW}^{\rm PBH}(f)
\simeq
\Omega_{\rm GW}^{\rm peak}
\left(\frac{f}{f_{\rm peak}^{\rm PBH}}\right)^{11/3}
\Theta\!\left(f_{\rm peak}^{\rm PBH}-f\right),
\label{eqn:omgw}
\end{equation}
where $\Omega_{\rm GW}^{\rm peak}$ denotes the maximum GW energy density fraction, given by
\begin{equation}
\Omega_{\rm GW}^{\rm PBH,\, peak}
\simeq
2\times10^{-6}
\left(\frac{\beta}{10^{-8}}\right)^{16/3}
\left(\frac{M_{\rm in}}{10^7\,{\rm g}}\right)^{34/9}.
\label{eqn:omgpeak}
\end{equation}

The above description relies on treating the PBH distribution as a continuous fluid. However, this approximation breaks down on scales smaller than the average separation between PBH. Consequently, the induced GW spectrum exhibits an ultraviolet cutoff, whose characteristic frequency corresponds to the comoving scale associated with the mean PBH separation. The resulting peak frequency can be estimated as
\begin{equation}
f^{\rm PBH,\, peak}
\simeq
1.7\times10^{3}\,{\rm Hz}
\left(\frac{M_{\rm in}}{10^4\,{\rm g}}\right)^{-5/6}.
\label{eqn:fpk}
\end{equation}

\begin{figure}[ht]
    \centering
    \includegraphics[width=0.65\linewidth]{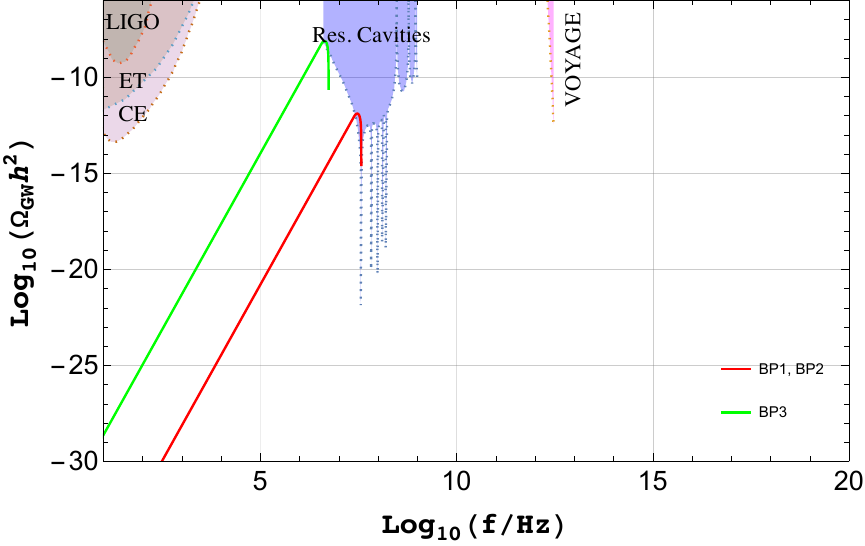}
    \caption{GW spectra sourced by PBH-induced scalar density perturbations for the benchmark points BP1, BP2, and BP3. Since BP1 and BP2 share the same PBH parameters, they yield identical spectra, shown by the red curve. The spectrum corresponding to BP3 is displayed in green.}
    \label{fig:GW_PBH_density}
\end{figure}

Fig. \ref{fig:GW_PBH_density} shows the GW from density fluctuations due to PBH. The spectra for BP1 and BP2 are exactly the same due to identical PBH parameters and are shown by the red curve.

\subsubsection{From PBH evaporation}

The production of gravitons from PBH evaporation has been investigated in several studies~\cite{Dolgov:2011cq, Dong:2015yjs, Hooper:2020evu, Perez-Gonzalez:2020vnz, Choi:2024acs}. In this work, we follow Ref.~\cite{Dolgov:2011cq} to estimate the GW spectrum generated from PBH evaporation.

The present-day GW energy density spectrum can be approximated as
\begin{eqnarray}
    \Omega^{\rm PBH,ev}_{\rm GW} h^2(f)
    \simeq
    1.36\times 10^{-21}
    \left(\frac{g_{*,H}}{100}\right)^2
    \left(\frac{1~{\rm g}}{M_{\rm in}}\right)^2
    \left(\frac{f}{10^9~{\rm Hz}}\right)^4
    I\!\left(\frac{2\pi f}{T_{\rm BH}^0}\right),
\end{eqnarray}
where $T_{\rm BH}^0$ denotes the PBH Hawking temperature redshifted to the present epoch and is given by
\begin{eqnarray}
    T_{\rm BH}^0
    \simeq
    4.53\times10^{15}~{\rm Hz}
    \left(\frac{100}{g_{*,s}(T_{\rm ev})}\right)^{1/12}
    \left(\frac{100}{g_{*,H}}\right)^{1/2}
    \left(\frac{M_{\rm in}}{10^5~{\rm g}}\right)^{1/2}.
\end{eqnarray}
The function $I$ is defined as
\begin{eqnarray}
    I\!\left(\frac{2\pi f}{T_{\rm BH}^0}\right)
    =
    \int_0^{z_{\rm max}}
    \frac{\sqrt{1+z}\, dz}
    {\exp\!\left[(1+z)\,2\pi f/T_{\rm BH}^0\right]-1},
\end{eqnarray}
where
\begin{eqnarray}
    1+z_{\rm max}
    \simeq
    7.36\times10^7
    \left(\frac{M_{\rm in}}{1~{\rm g}}\right)^{4/3}
    \beta^{1/3}.
\end{eqnarray}

\begin{figure}[ht]
    \centering
    \includegraphics[width=0.65\linewidth]{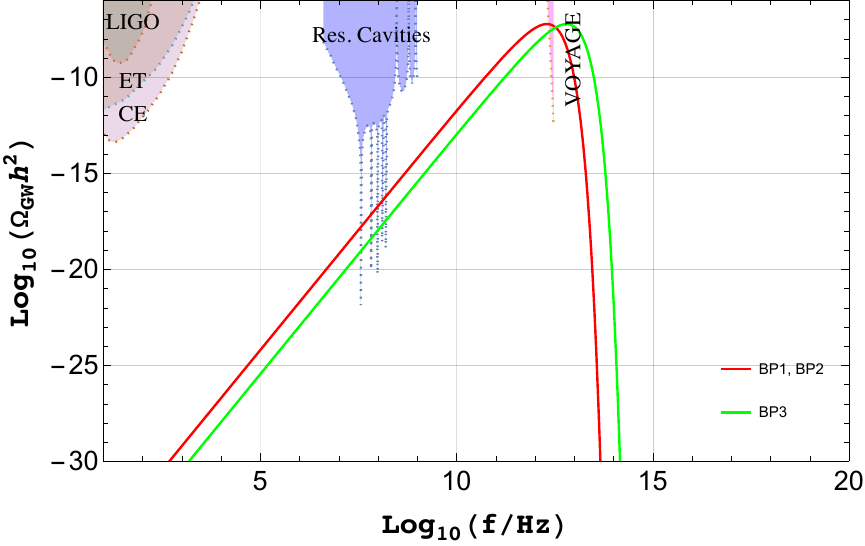}
    \caption{GW spectra generated from graviton emission during PBH evaporation. Similar to Fig. \ref{fig:GW_PBH_density}, BP1 and BP2 yield identical spectra, shown by the red curve. The spectrum corresponding to BP3 is displayed in green.}
    \label{fig:GW_PBH_eva}
\end{figure}

Fig.~\ref{fig:GW_PBH_eva} shows the GW spectrum arising from PBH evaporation. Since the PBH parameters are identical for BP1 and BP2, the corresponding GW spectra completely overlap and are represented by the red curve.

\begin{figure}[ht]
    \centering
    \includegraphics[width=0.65\linewidth]{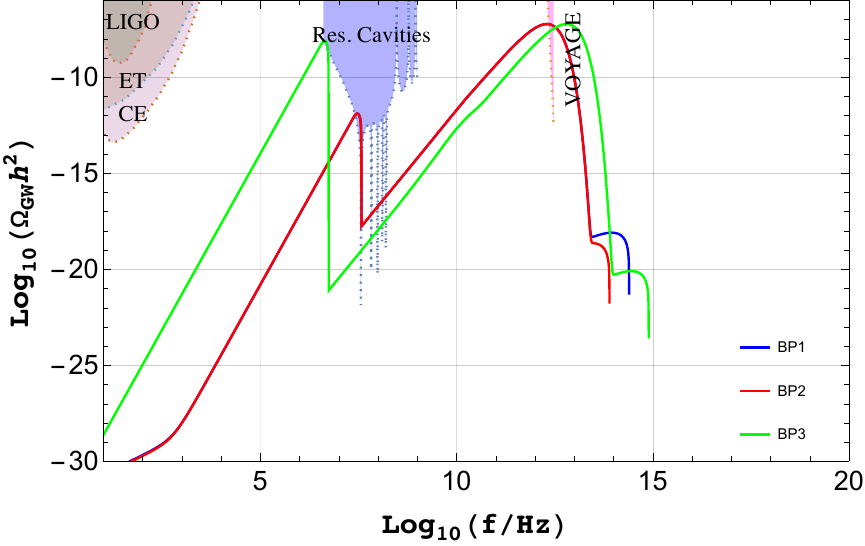}
    \caption{Total GW spectrum, including all relevant contributions, for the hierarchical leptogenesis benchmark points listed in Table~\ref{tab:BP}.}
    \label{fig:GW_total}
\end{figure}

Finally, in Fig.~\ref{fig:GW_total} we present the total GW spectrum obtained by combining all relevant sources. At low frequencies, the signal is dominated by GW induced by enhanced density perturbations. In the intermediate frequency range, $10^{10}\,{\rm Hz} \lesssim f \lesssim 10^{14}\,{\rm Hz}$, two main contributions arise: the CGMB and graviton emission from PBH evaporation. Among these, the latter provides the dominant contribution and is clearly visible in the figure. At even higher frequencies, around $f \sim 10^{15}\,{\rm Hz}$, the GW signal produced through RHN bremsstrahlung becomes dominant, carrying characteristic signatures of high-scale leptogenesis.

We note from Fig.~\ref{fig:GW_total} that, for the benchmark points listed in Table~\ref{tab:BP}, the GW spectrum arising from RHN decay in hierarchical leptogenesis contributes significantly only in the ultra-high-frequency range, \( \sim 10^{14}-10^{15}~\mathrm{Hz} \). In this regime, distinguishing the GW spectrum from RHN decay may be challenging for some of the benchmark points due to the presence of other dominant GW sources, particularly the contribution from graviton emission during PBH evaporation. This is even more difficult for most of the benchmark points in case of thermal high-scale leptogenesis without a PBH-dominated epoch, as shown in Appendix \ref{App:B}. From Eq.~\eqref{eq:peak_f}, it is evident that the peak frequency, and consequently the peak amplitude, can be enhanced by considering a smaller Yukawa coupling for \(N_1\). In the framework of hierarchical leptogenesis, however, reducing the Yukawa coupling simultaneously suppresses the CP asymmetry parameter, making it difficult to generate the observed baryon asymmetry of the Universe. This limitation can be circumvented in resonant leptogenesis, where the Yukawa couplings and the CP asymmetry parameter become effectively decoupled. As a result, one can consider significantly smaller Yukawa couplings while still obtaining the required baryon asymmetry through resonant enhancement.

\begin{figure}[ht]
    \centering
    \includegraphics[width=0.65\linewidth]{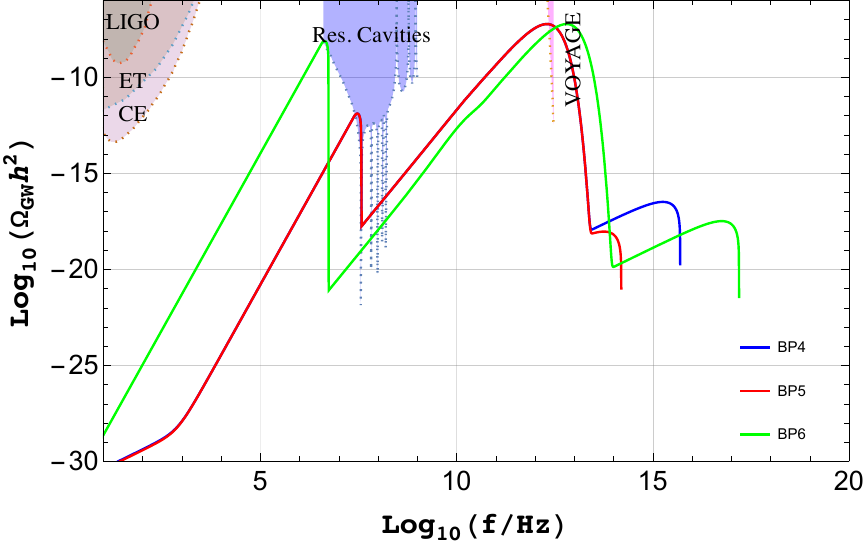}
    \caption{Total GW spectrum, including all relevant contributions, for the resonant leptogenesis benchmark points listed in Table~\ref{tab:BP_reso}.}
    \label{fig:GW_total_R}
\end{figure}

To illustrate this possibility, we present three benchmark points for resonant leptogenesis in Table~\ref{tab:BP_reso}, featuring smaller Yukawa couplings than those in Table~\ref{tab:BP}. Here, \(\Delta m\) denotes the mass splitting between \(m_{N_2}\) and \(m_{N_1}\), while the mass of \(N_1\) and the remaining PBH parameters are kept identical to those in Table~\ref{tab:BP}. The resulting GW spectra are shown in Fig.~\ref{fig:GW_total_R}. In contrast to the hierarchical leptogenesis case displayed in Fig.~\ref{fig:GW_total}, the RHN bremsstrahlung contribution extends over a much broader frequency range and constitutes a substantial component of the total GW spectrum, thereby significantly improving its prospects for observational distinguishability.

\begin{table}[htb!]
\centering
\renewcommand{\arraystretch}{1.2}
\begin{tabular}{|c|c|c|c|c|c|c|c|}
\hline
\text{Benchmark} & ${M_{\rm in} \, \text{(g)}}$  & ${\beta}$ &
${m_{N_1}} \, \text{(GeV)}$ & ${\Delta m} \, \text{(GeV)}$ &
${m_{N_3}} \, \text{(GeV)}$ & ${\sqrt{(Y^{\dagger}_{\nu}Y_{\nu})_{11}}}$& {$\varepsilon^{\rm CP}$} \\
\hline
BP4 &  0.1 & $5\times10^{-4}$ & $10^{15}$ & $5\times10^{9}$  & $10^{16}$ &  $5\times10^{-7}$& {$10^{-3}$} \\
\hline
BP5 &  0.1 & $5\times10^{-4}$ & $10^{14}$ & $6\times10^{10}$  & $10^{16}$ &  $5\times10^{-6}$& {$10^{-5}$}  \\
\hline
BP6 & 1 & $5\times10^{-4}$ & $10^{14}$ & $4\times10^{5}$  & $10^{16}$ &  $5\times10^{-9}$& {$3\times10^{-3}$}  \\
\hline
\end{tabular}
\caption{Benchmark points for resonant leptogenesis.}
\label{tab:BP_reso}
\end{table}

\begin{figure}
    \centering
    \includegraphics[width=0.65\linewidth]{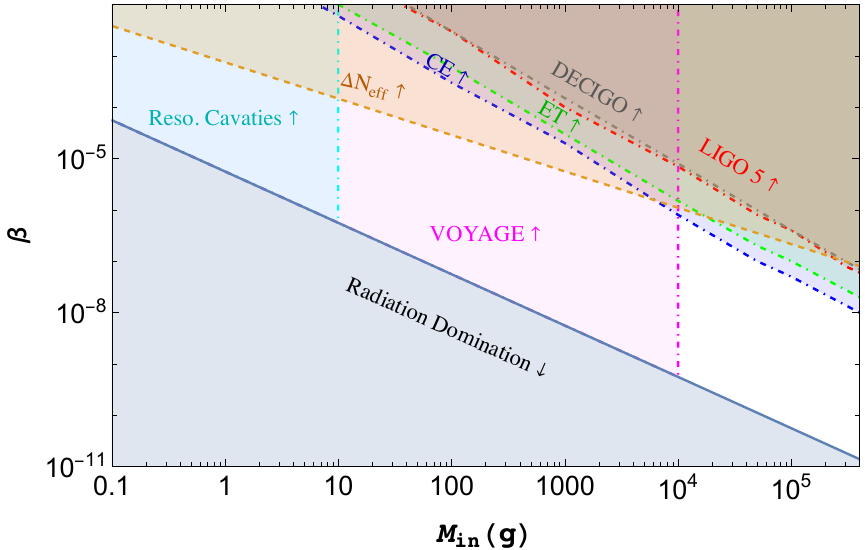}
    \caption{{Constraints and projected future sensitivities of various GW experiments in the $M_{\rm in}$--$\beta$ plane, incorporating all relevant GW contributions. The regions below the solid blue line and above the dashed orange line represent the radiation domination and $N_{\rm eff}$ constraints, respectively, while the regions bounded by dot-dashed lines denote the sensitivities of future GW experiments.}}
    \label{fig:scan3}
\end{figure}

{We now present a comprehensive parameter-space analysis of the detectability of GWs in the PBH-assisted leptogenesis scenario. In the PBH-dominated regime, the GW contributions from PBH density fluctuations, graviton emission, and the CGMB depend primarily on the PBH parameter space, characterized by the initial PBH mass, $M_{\rm in}$, and the initial PBH energy fraction, $\beta$. In contrast, the gravitational bremsstrahlung contribution depends primarily on $M_{\rm in}$, the mass of the lightest RHN, $m_{N_1}$, and the neutrino Yukawa coupling, $Y_\nu$. In Figs.~\ref{fig:scan1} and \ref{fig:scan2}, we show the peak amplitudes and corresponding peak frequencies of the gravitational bremsstrahlung spectra in the $M_{\rm in}$--$m_{N_1}$ plane for hierarchical and resonant leptogenesis, respectively. We find that, in both scenarios, the predicted GW signals remain beyond the reach of future GW detectors. In Fig.~\ref{fig:scan3}, we present the existing GW constraints and projected future sensitivities of various GW experiments in the $M_{\rm in}$--$\beta$ plane, incorporating all the relevant GW contributions. For this analysis, we fix the RHN mass to $m_{N_1}=10^{15}\,\mathrm{GeV}$ and the neutrino Yukawa coupling combination to $\sqrt{(Y_\nu^\dagger Y_\nu)_{11}}=10^{-7}$.  The region below the solid blue curve corresponds to parameter choices for which PBHs do not dominate the energy density of the Universe before evaporating and is therefore incompatible with the assumed PBH-dominated leptogenesis scenario. The region above the dashed orange curve is excluded by constraints on the \textit{effective number of relativistic species}, $N_{\rm eff}$, arising from GWs sourced by PBH density fluctuations \cite{Domenech:2020ssp}. The dot-dashed curves indicate the projected sensitivities of various future GW experiments. The sensitivities of LIGO-5, ET, CE, and DECIGO arise primarily from the GW contribution sourced by PBH density fluctuations, whereas those of resonant cavities and VOYAGE are driven predominantly by Hawking graviton emission from PBHs. We emphasize that the gravitational bremsstrahlung signal associated with successful leptogenesis remains below the projected sensitivities of all the experiments considered. Consequently, varying the RHN mass and neutrino Yukawa coupling within the parameter space consistent with successful leptogenesis does not qualitatively change the detectability prospects of this contribution. Moreover, in Fig. \ref{fig:scan3}, we consider the PBH mass in the range $0.1 - 4\times10^{5}$ g to show the detectability of both hierarchical and resonant leptogenesis scenarios. }

Before concluding, we would like add a remark that the energy density of RHNs redshifts more slowly than that of the radiation bath after their production from PBH evaporation. Consequently, for sufficiently small Yukawa couplings, the RHNs can eventually dominate the energy density of the Universe before decaying. In this work, however, we restrict our analysis to benchmark points for which the RHNs remain subdominant throughout their evolution. If an RHN-dominated era were to occur, both the dynamics of PBH-assisted leptogenesis and the resulting GW spectrum would be modified. A dedicated investigation of this scenario is left for future study.

\section{Conclusion}
\label{sec3}
We have revisited the graviton bremsstrahlung signature of leptogenesis by considering a non-thermal production of heavy right-handed neutrinos from evaporation of ultra-light primordial black holes. As it requires very heavy RHNs close to the GUT scale for the corresponding GW spectrum to lie in the most optimistic regime of future verifications, non-thermal origin of RHNs avoid the strong washout issues of thermal leptogenesis at very high scale. In addition, the PBH-dominated era also keeps this GW spectrum distinct from the one produced by the thermal background keeping future experimental verification more plausible. We compute the graviton bremsstrahlung contribution to the GW spectra in this setup and compare it with the GW spectrum from thermal background by considering a few benchmark choices of PBH and model parameters consistent with light neutrino mass and successful leptogenesis. Compared to the other avenues of non-thermal leptogenesis, the presence of ultra-light PBH also brings two additional sources of GW due to density perturbations and graviton emission via Hawking evaporation. The combined GW spectrum including all these sources bears a multi-peak feature at frequencies. For resonantly enhanced CP asymmetry, the spectrum due to graviton bremsstrahlung becomes more prominent and further separated from the thermal contribution compared to the hierarchical leptogenesis scenario. While most of the near future GW prospects lie in the low frequency regime ($f \lesssim \mathcal{O}(\rm kHz)$), our work shows one possible example of high-frequency GW with unique features and connected to the origin of light neutrino mass and baryon asymmetry of the Universe. More avenues of high-frequency GW detection should be able to verify these scenarios to a greater accuracy.

\acknowledgements
The work of D.B. is supported by the Science and Engineering Research Board (SERB), Government of India grant CRG/2022/000603. 

\appendix

\section{Details of the benchmark points} \label{App:A}
The Yukawa coupling matrix $Y_\nu$ in Eq. \eqref{eq:10} can be conveniently expressed using the Casas--Ibarra parametrization~\cite{Casas:2001sr},
\begin{align}\label{eq:Ynu}
Y_{\nu}=-i\,\frac{\sqrt{2}}{v}\,\mathcal{U}\,
D_{\sqrt{m}}\,\mathcal{R}^{T}\,D_{\sqrt{M}}\,,
\end{align}
where $\mathcal{U}$ is the Pontecorvo--Maki--Nakagawa--Sakata (PMNS) matrix~\cite{ParticleDataGroup:2024cfk}, which relates the flavour and mass eigenstates of the light neutrinos. The diagonal matrices
\begin{align}
D_{\sqrt{m}}
&=\text{diag}\!\left(\sqrt{m_1},\,\sqrt{m_2},\,\sqrt{m_3}\right),\\
D_{\sqrt{M}}
&=\text{diag}\!\left(\sqrt{m_{N_{1}}},\,\sqrt{m_{N_{2}}},\,\sqrt{m_{N_{3}}}\right),
\end{align}
contain the square roots of the light-neutrino and RHN masses, respectively, where the RHNs are assumed to be in a mass-diagonal basis. The matrix $\mathcal{R}$ is a complex orthogonal matrix satisfying $\mathcal{R}^{T}\mathcal{R}=\mathbb{I}$ and can be parametrised in terms of three complex angles $\omega_{1,2,3}$. 

Below we list the details of the benchmark points BP1, BP2, BP3 used in our analysis.
\begin{enumerate}
    \item BP1 : $M_{\rm in}=0.1$ g, $\beta=5\times 10^{-4}$, $m_{N_1} = 10^{15}$ GeV, $m_{N_2} = 7\times10^{15}$ GeV, $m_{N_3} = 10^{16}$ GeV, $m_{1}=10^{-10}$ eV, $\omega_{1} = 0$, $\omega_{2}= 0.00056 - i\,0.00039$, $\omega_{3} = 0.0055 - i\,0.0011$.\\
The corresponding Yukawa matrix is given by\\
    \small
    $Y_{\nu} = \begin{pmatrix}
1.59\times10^{-3} + 4.31\times10^{-4}\,i &
7.89\times10^{-1} - 4.91\times10^{-6}\,i &
-6.11\times10^{-1} + 1.56\times10^{-1}\,i \\[2mm]

2.46\times10^{-3} + 1.43\times10^{-4}\,i &
9.85\times10^{-1} + 1.91\times10^{-2}\,i &
2.49 + 2.23\times10^{-6}\,i \\[2mm]

-7.32\times10^{-4} - 5.97\times10^{-4}\,i &
-6.28\times10^{-1} + 2.41\times10^{-2}\,i &
3.14 - 5.32\times10^{-7}\,i
\end{pmatrix},$\\
with $\sqrt{(Y^{\dagger}_{\nu}Y_{\nu})_{11}} = 10^{-5}$. The CP asymmetry parameter is
${\varepsilon^{\rm CP}_{1}} =0.001$.

\item BP2 : $M_{\rm in}=0.1$ g, $\beta=5\times10^{-4}$, $m_{N_1} = 10^{14}$ GeV, $m_{N_2} = 7\times10^{15}$ GeV, $m_{N_3} = 10^{16}$ GeV, $m_{1}=10^{-6}$ eV, $\omega_{1} = 0$, $\omega_{2}= 0.004 - i\,10^{-4}$, $\omega_{3} = 0.012 + i\,10^{-3}$.\\
The corresponding Yukawa matrix is given by\\
    \small
    $Y_{\nu} = \begin{pmatrix}
2.36\times10^{-3} + 1.63\times10^{-4}\,i &
7.89\times10^{-1} - 2.18\times10^{-5}\,i &
-6.11\times10^{-1} + 1.56\times10^{-1}\,i \\[2mm]

1.74\times10^{-3} + 1.56\times10^{-4}\,i &
9.85\times10^{-1} + 1.91\times10^{-2}\,i &
2.49 - 5.50\times10^{-6}\,i \\[2mm]

1.17\times10^{-3} - 2.65\times10^{-5}\,i &
-6.28\times10^{-1} + 2.41\times10^{-2}\,i &
3.14 + 9.74\times10^{-7}\,i
\end{pmatrix},$\\
with $\sqrt{(Y^{\dagger}_{\nu}Y_{\nu})_{11}} = 10^{-5}$. The CP asymmetry parameter is
${\varepsilon^{\rm CP}_{1}}$.

\item BP3 : $M_{\rm in}=1$ g, $\beta=5\times10^{-4}$, $m_{N_1} = 10^{14}$ GeV, $m_{N_2} = 7\times10^{15}$ GeV, $m_{N_3} = 10^{16}$ GeV, $m_{1}=10^{-6}$ eV, $\omega_{1} = 0$, $\omega_{2}= 0.004 - i\,10^{-4}$, $\omega_{3} = 0.012 + i\,10^{-3}$.\\
The corresponding Yukawa matrix is given by\\
    \small
    $Y_{\nu} = \begin{pmatrix}
3.54\times10^{-4} + 4.26\times10^{-4}\,i &
7.89\times10^{-1} - 1.80\times10^{-5}\,i &
-6.11\times10^{-1} + 1.56\times10^{-1}\,i \\[2mm]

5.24\times10^{-4} + 8.48\times10^{-4}\,i &
9.85\times10^{-1} + 1.91\times10^{-2}\,i &
2.49 - 1.12\times10^{-5}\,i \\[2mm]

1.56\times10^{-4} - 4.74\times10^{-5}\,i &
-6.28\times10^{-1} + 2.41\times10^{-2}\,i &
3.14 + 2.05\times10^{-6}\,i
\end{pmatrix},$\\
with $\sqrt{(Y^{\dagger}_{\nu}Y_{\nu})_{11}} = 10^{-6}$. The CP asymmetry parameter is
${\varepsilon^{\rm CP}_{1}} =3\times10^{-3}$.
\end{enumerate}

\begin{figure}
    \centering
    \includegraphics[width=0.8\linewidth]{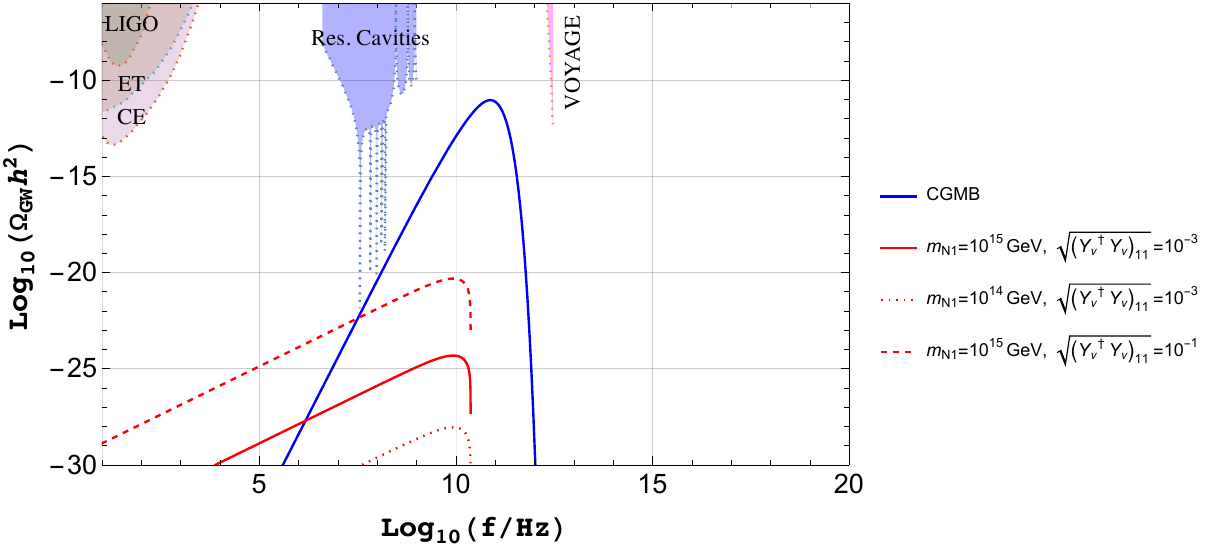}
    \caption{GW spectra from graviton bremsstrahlung via the decay of thermal RHN, shown for three choices of RHN masses and couplings, in the absence of any PBH-dominated epoch. For comparison, the CGMB is shown by the blue curve, assuming the maximum allowed reheating temperature, ($T_{\rm RH}=10^{16}$ GeV). No PBH are assumed in this analysis.}
    \label{fig:thermal}
\end{figure}

\section{GW spectra in the absence of PBH}
\label{App:B}
Fig. \ref{fig:thermal} shows the graviton bremsstrahlung and CGMB spectra for a few benchmark points without considering a PBH-dominated phase. As mentioned before, for thermally produced RHNs, the peak of the RHN-generated graviton bremsstrahlung spectra remains underneath the one generated by the thermal background.

\bibliographystyle{JHEP}
\bibliography{ref1} 

\end{document}